\documentclass{aastex631}

\usepackage{fancyvrb}

\newcommand{\teff}{T$_{\mathrm{eff}}$}
\newcommand{\rvar}{\textit{R}$_{\mathrm{var}}$}
\usepackage{svg}
\usepackage{tikz}
\usetikzlibrary{shapes.geometric, arrows}

\shorttitle{Stellar rotation in TESS}
\shortauthors{Colman et al.}

\accepted{2024-02-16}

\begin{document}

\title{Methods for the detection of stellar rotation periods in individual TESS sectors and results from the Prime mission}



\correspondingauthor{Isabel L. Colman}
\email{icolman@amnh.org}

\author[0000-0001-8196-516X]{Isabel L. Colman}
\affiliation{Department of Astrophysics, American Museum of Natural History, 200 Central Park West, New York, NY 10024, USA}

\author[0000-0003-4540-5661]{Ruth Angus}
\affiliation{Department of Astrophysics, American Museum of Natural History, 200 Central Park West, New York, NY 10024, USA}
\affiliation{Center for Computational Astrophysics, Flatiron Institute, 162 5th Avenue, New York, NY 10010, USA}

\author[0000-0001-6534-6246]{Trevor David}
\affiliation{Center for Computational Astrophysics, Flatiron Institute, 162 5th Avenue, New York, NY 10010, USA}
\affiliation{Department of Astrophysics, American Museum of Natural History, 200 Central Park West, New York, NY 10024, USA}

\author[0000-0002-2792-134X]{Jason Curtis}
\affiliation{Department of Astronomy, Columbia University, 550 West 120th Street, New York, NY 10027, USA}

\author[0000-0002-0842-863X]{Soichiro Hattori}
\affiliation{Department of Astronomy, Columbia University, 550 West 120th Street, New York, NY 10027, USA}
\affiliation{Department of Astrophysics, American Museum of Natural History, 200 Central Park West, New York, NY 10024, USA}

\author[0000-0003-4769-3273]{Yuxi (Lucy) Lu}
\affiliation{Department of Astrophysics, American Museum of Natural History, 200 Central Park West, New York, NY 10024, USA}



\begin{abstract}
For ongoing studies of the role of rotation in stellar evolution, we require large catalogs of rotation periods for testing and refining gyrochronology. While there is a wealth of data from the Kepler and K2 missions, TESS presents both an opportunity and a challenge: despite its all-sky coverage, rotation periods remain hard to detect. We analyzed individual TESS sectors to detect short-period stellar rotation, using only parameters measured from light curves for a robust and unbiased method of evaluating detections. We used random forest classifiers for vetting, trained on a large corpus of period measurements in KELT data from the \citet{oelkers_variability_2018} catalog and using TESS full-frame image light curves generated by \textit{eleanor} \citep{feinstein_eleanor_2019}. Finally, using data from the first 26 sectors of TESS, we analyzed 432,704 2-minute cadence single-sector light curves for FGKM dwarfs. We detected 16,800 periods in individual sector light curves, covering 10,909 distinct targets, and we present a catalog of the median period for each target as measured by a Lomb-Scargle periodogram.
\end{abstract}

\keywords{Stellar rotation (1629) --- Period determination (1211) --- Astronomy data analysis (1858) --- Random forests (1935)}


\section{Introduction} \label{sec:intro}

An understanding of stellar rotation is key to the study of stellar evolution, due to the phenomenon of spin down: the process by which a star gradually rotates more slowly over the course of its main sequence lifetime. The process of determining a precise main sequence stellar age from a star's rotation period is known as gyrochronology \citep{barnes_rotational_2003}. Since the first observation of a power law relation between stellar spin down and age \citep{skumanich_time_1972} and the earliest detections to support this from ground-based rotational velocity studies \citep[e.g.][]{smith_rotational_1979,vaughan_survey_1980,vogel_rotational_1981}, the question of defining gyrochronological relations has only become more complex \citep{angus_toward_2019,bouma_empirical_2023}. To deepen our understanding of these phenomena, we need as many measurements of stellar rotation as possible, from a sample that includes main sequence stars across a wide range of masses, metallicities, and ages. In the spirit of pursuing this goal, our introduction surveys the lineage of the study of stellar rotation using space-based data, to situate our work in its broader context and provide motivation for the refinement of detection, classification, and characterization methods that is undertaken in the larger part of this paper. The Kepler mission \citep{borucki_kepler_2010} represented a massive increase in the volume of available rotation periods, with its high quality time series data spanning four years of continuous observation \citep[e.g.][]{mcquillan_rotation_2014, santos_surface_2021}. Rotation is measured in time series data using methods familiar from the detection of exoplanets and other variability: periodicity appears in light curves due to star spots on the surface, dark regions of low magnetic activity, which cause dips in brightness as they pass into our line of sight due to stellar rotation. For the first time with Kepler, the community has been enabled to undertake large-scale studies of stellar rotation and produce catalogs of tens of thousands of measurements, work which continued with Kepler's successor K2 (\citet{howell_k2_2014}; rotation period catalogs include \citet{esselstein_k2_2018,reinhold_stellar_2020,gordon_stellar_2021}).

While there is yet more to be discovered in data from the Kepler and K2 missions, we turn our attention to their successor, the Transiting Exoplanet Survey Satellite \citep[TESS,][]{ricker_transiting_2015}. This ongoing, whole-sky survey presents a vital opportunity for large-scale rotation studies. Before further discussion of our motivation and the legacy of prior rotation studies, we will briefly introduce several major methods of analysis used to detect rotation --- and, more broadly, periodicity --- in time series data.

Perhaps the most common tool for period-space analysis of stellar variability, the Lomb-Scargle (LS) periodogram is an algorithm that iteratively fits sinusoids in frequency space. It is similar to the (fast) Fourier transform, but has the advantage that by construction it does not assume equal spacing between data points; thus, it can be used on data with gaps, which is the case for the vast majority of space- and ground-based photometric data. Kepler and TESS data both have multiple types of gaps within quarters/sectors, due to issues as varied as telescope operation errors and cosmic ray incidences, and TESS has gaps inherent to each sector in the halfway data downlink. In data from both missions, the LS method has proved effective and efficient for detecting periodic signal. Another method which we employed in this work is the autocorrelation function (ACF) of a time series, which is produced by correlating the series with a sequence of time-delayed copies of itself. The value of the ACF is higher when a pair of series are more highly correlated --- i.e., if the time lag on the delayed copy is a multiple of a periodic signal present in the data. When plotted against the time lags of the copies, the first peak in the ACF will typically correspond to the value of the strongest periodic signal. The ACF will be non-periodic (noisy) when there is no clear signal present, as there will be no highly-correlated series pairs.

Another common method, which we did not use in this work, is the wavelet transform \citep[see][]{breton_rooster_2021, claytor_tess_2023}. This can be thought of as a sort of Fourier transform that provides measurements of time-dependent signal intensity. Instead of fitting sinusoids, this transform fits a series of non-continuous oscillatory signals, known as wavelets. The result is a three-dimensional grid of signal amplitudes in period-time space, which is useful for determining if a signal degrades or increases in amplitude over time, such as in the case of star spot decay. Finally, Gaussian processes (GPs) are useful for retrieving rotation periods, particularly from non-sinusoidal signals \citep{angus_inferring_2018, gordon_stellar_2021}. GPs are a machine learning tool for regression problems, based on the idea that a multivariate Gaussians distribution can be used as a Bayesian prior for random-sample modeling. We also did not use GPs in this work. Wavelet transforms and GPs are both well-suited for the search for longer, more complex periods, which are less likely to be present in single-sector data.

Using simulated data, \citet{aigrain_testing_2015} showed that LS periodograms, ACFs, wavelet transforms, and modeling all provide accurate rotation periods from Kepler, though all methods of detection require additional vetting processes. Although there has not yet been a comparable test of these methods published for TESS data, various studies have shown that there are differences in application when it comes to TESS. In Section~\ref{sec:methods1}, we will introduce the LS and ACF methods in context of our pipeline, and in Section~\ref{sec:methods2} we will provide an overview of other major pipelines which use these methods.

Early work analyzing rotation in Kepler showed a bimodal distribution in period-temperature space, revealing a dearth of periods along a sequence around $\sim$11 days, known as the ``intermediate period gap'' \citep{mcquillan_measuring_2013}, and evidence of differences in rotation-age relations for stars of different masses and evolutionary stages \citep{garcia_rotation_2014}. The study of these challenges to models in gyrochronology has been a key feature of many ensuing catalog studies and related scholarship, particularly the search for an astrophysical or observational explanation for the intermediate period gap. \citet{mcquillan_rotation_2014} (hereafter M14) presented the first major catalog of stellar rotation from Kepler. (We will return to this study in further detail in Section~\ref{sec:methods2}.) The M14 catalog covers the analysis of 133,030 stars, which yielded 34,030 detections, setting a benchmark for the detection and analysis of rotation periods from a large volume of space-based data. Many papers since have used the data from M14 and validated these results: \citet{reinhold_rotation_2015} reanalyzed 24,124 targets from the M14 catalog and found 77\% agreement in their results, a promising early result for the integrity of rotation detection and analysis techniques when applied to large datasets.

After M14, the next major catalog of Kepler rotation periods released was the \citet{santos_surface_2021} catalog, hereafter S21. They presented 55,232 rotation periods for main sequence stars and some subgiants, an expansion on the M14 sample. S21 used the ROOSTER pipeline \citep{breton_rooster_2021}, which uses ACF and wavelet techniques to detect and measure rotation periods. ROOSTER is one of the pipelines we will discuss in detail in Section~\ref{sec:methods2}, alongside the S21 catalog. Finally, we note the recent publication of a catalog of 67,163 rotation periods by \citet{reinhold_new_2023}, which introduces a new method for period detection based on the gradient of the power spectrum combined with an ACF, well-suited for detecting short-term quasi-irregular modulation.

There are multiple smaller catalogs of Kepler and K2 rotation periods, many of which focus on specific subsets of the data (such as open clusters) in depth. \citet{reinhold_stellar_2020} presented a catalog of 29,860 periods from K2, confirming that the intermediate period gap is not confined to stars in the Kepler field of view. This is also confirmed by \citet{gordon_stellar_2021}, who used Gaussian processes to retrieve 8,977 rotators in K2. \citet{curtis_when_2020} looked at main sequence stars in the open cluster Ruprecht~147 using data from K2 and the Palomar observatory, and found an isochronal sequence that moved across the period gap, suggesting that the gap has its origins in astrophysical processes and did not represent different epochs of star formation, as had previously been suggested. This result was reinforced by \citet{gruner_rotation_2020}, who observed the same sequence for Ruprecht~147 using K2 data. Recent work by \citet{lu_bridging_2022} recovers $\sim$40,000 rotation periods using data from the Zwicky Transient Facility, also noting that stars seem to ``cross'' the gap at certain ages. Large rotation catalogs have been crucial in confirming that the gap is due to an astronomical process; the predominant theory of weakened magnetic braking suggests that stars undergo a period of stalled spin-down, followed by an accelerated increase in spin down, leading them to rapidly cross the gap \citep{vansaders_weakened_2016,hall_weakened_2021}. In complementary work, a recent spectroscopic study of targets from the S21 catalog by \citet{david_further_2022} shows that there are pile-ups at the long- and short-period edges of the distribution in temperature-period space, and agrees that the cause of the gap is likely intrinsic to main sequence stellar evolution. The physical processes behind the intermediate period gap are a major result from large-scale space-based studies, and provides motivation for further probing of the demographics of stellar rotation.


The Kepler corpus of stellar rotation analysis has provided a breadth of new results, but there are still unanswered questions. With the advent of the TESS mission, we have ongoing access to unprecedented amounts of data. However, the detection of periodicity in TESS data is limited by the sector length. With $\sim$27~d sectors and systematics introduced by the data downlink gap at 13 days, the detection of periods longer than 12--15~d presents a challenge \citep[e.g.][]{claytor_recovery_2022, holcomb_spinspotter_2022, kounkel_untangling_2022, fetherolf_variability_2022}. To solve this problem, there are many new tools being created for TESS photometry in the full frame images (FFIs), with an eye towards optimizing light curves for variability detection, especially rotation: \citet{hattori_unpopular_2021} introduced the \textit{unpopular} code, which uses causal pixel modeling to detrend and stitch TESS light curves, preserving signals on the scale of a TESS sector. Similarly, \textit{TESS-SIP} \citep{hedges_systematics-insensitive_2020} uses a modified periodogram that is systematics-insensitive to TESS data idiosyncrasies and enables simple light-curve stitching, enabling the detection of longer periods, particularly in the continuous viewing zones.

Early supplementary work for TESS data has included the expansion of rotation period catalogs. The \citet{oelkers_variability_2018} catalog is one major publication, presenting 62,229 high-confidence rotation periods in TESS targets discovered using photometry from Kilodegree Extremely Little Telescope \citep[KELT,][]{pepper_kilodegree_2007}. There has also been work done to search for rotation in stellar clusters \citep{healy_stellar_2020} and associations \citep{kounkel_untangling_2022} in the TESS field, capitalizing on the all-sky coverage to increase the scope of rotation studies in same-age associations. Other early TESS studies have focused on TESS objects of interest (TOIs): \citet{canto_martins_search_2020} used the LS periodogram, wavelet transform, and visual inspection to detect 163 rotation periods among 1,000 TOIs. Recent tools for rotation period detection in TESS data include the neural network introduced by \citet{claytor_recovery_2022}, who showed that they can infer highly reliable rotation periods from synthetic TESS light curves, which were created by convolution with real TESS data. They then applied this to TESS Southern Continuous Viewing Zone targets \citep{claytor_tess_2023} to retrieve a catalog of 7,971 rotation periods. Their method uses a wavelet transform for detection of periodicity, but the final period is derived from the neural network. \citet{holcomb_spinspotter_2022} used the ACF to retrieve rotation periods in TESS data, using machine learning to vet detections based on features of the ACF itself. We will discuss the \citet{holcomb_spinspotter_2022} pipeline in further detail in Section~\ref{sec:methods2}, as their study is in many ways comparable to ours.

It is not in spite of, but rather complementary to this growing corpus of work regarding the detection of rotation periods in TESS data that we present the following discussion of methodologies and the associated rotation catalog. Our work represents a thorough test of several major methods of detecting rotation periods via the presence of spot modulation, bolstered by novel applications of machine learning and select visual inspection at some stages of the process. We set aside the period-limiting factors by focusing our search on shorter rotation periods. The problem of searching for rotation periods in TESS then becomes twofold, based on the enormous amounts of data available: we need to search this large volume of data efficiently, and vet detections effectively at a scale where visual inspection is untenable. We introduce a software pipeline\footnote{\url{https://github.com/astrobel/spinneret}} written in Python that is designed to search for periods under the length of a TESS sector, using the two-minute cadence light curves produced by the TESS Science Processing Operations Center \citep{jenkins_tess_2016}. In the following sections, we will cover the construction of the pipeline, compare it to other rotation detection pipelines, cover the process and results of training the machine learning classifiers, and finally present a catalog of our findings from the first 26 sectors of the TESS mission, with the hope that these data can be of use in further demographic studies of stellar rotation.

\section{Methods} \label{sec:methods}

\subsection{Pipeline construction} \label{sec:methods1}

Our pipeline is specifically designed to detect short stellar rotation periods in single TESS sectors. In this context, we define ``short'' as $<$30~d, though practically the pipeline excels at detecting periods $<$12~d, which is consistent with other surveys of TESS data. As we will show in Section~\ref{sec:results}, the shortest periods we are able to detect are on order 0.5~days. The pipeline obtains light curve features from a TESS sector, including a period measurement, and vets the measurements in two stages: Classifier \#1 tells us whether or not we have detected rotation, and Classifier \#2 determines whether or not the measured period is accurate. We will discuss our definition of accuracy and the training data further in Section~\ref{sec:methods3}.

Each stage of vetting is performed using a random forest algorithm \citep{breiman_random_2001}. The random forest can be used for regression or classification problems. In both stages of our algorithm, we use the random forest classifier from \textit{scikit-learn} \citep{pedregosa_scikit-learn_2011}, which essentially converts all input parameters into a tree structure of decisions and iterates through options to sort inputs into a discrete number of output states. The random forest classifier is useful for a multi-parameter problem which requires high-speed calculation. Additionally, each classifier only needs be constructed once, and can then be exported using the Python module \textit{joblib} for fast loading. As described above, we use binary states for each stage of classification: detection vs non-detection, and accurate vs inaccurate measurement. (We will use this terminology throughout the paper to discuss pipeline classification.) The \textit{scikit-learn} random forest allows us to specify the number of decision trees, or ``estimators,'' which are considered in the classification problem. Above a certain number of estimators, the random forest classifier will converge on a distribution of binary output states. However, more estimators may also make the classifier less permissive as a result of overfitting: given this, we aim to optimize the number of estimators to produce the best ratio of true positives to true negatives, which can be tested on a validation dataset. Optimizing these numbers in the training phase of pipeline construction leads to more reliable results when the pipeline is applied blind to real TESS data. We will describe the process of classifier optimization in Section~\ref{sec:methods3}.

\tikzstyle{process} = [rectangle, minimum width=2cm, minimum height=1cm, text centered, fill=cyan!30]
\tikzstyle{decision} = [diamond, minimum width=3cm, minimum height=1cm, text centered, fill=yellow!40]
\tikzstyle{no} = [rectangle, rounded corners, minimum width=2cm, minimum height=1cm, text centered, fill=red!30]
\tikzstyle{yes} = [rectangle, rounded corners, minimum width=2cm, minimum height=1cm, text centered, fill=green!40]
\tikzstyle{arrow} = [thick,->,>=stealth]
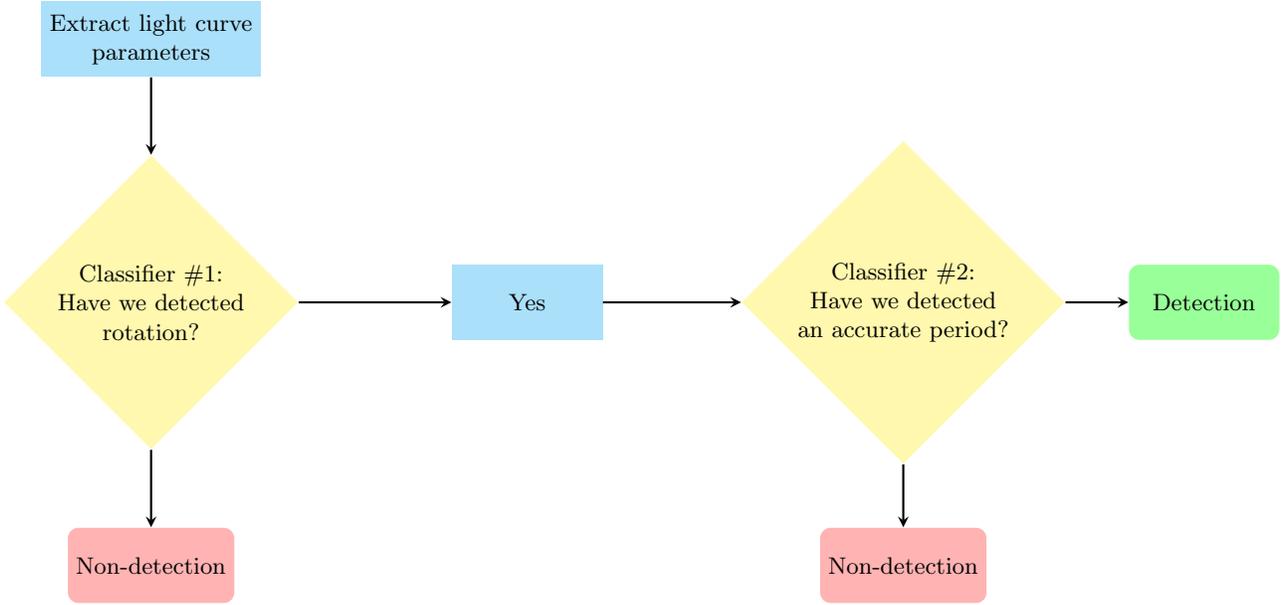
\begin{figure}
\centering
\begin{tikzpicture}
    \node (pro1) [process, align=center] {Extract light curve \\ parameters};
    \node (dec1) [decision, align=center, below of=pro1, yshift=-2.5cm] {Classifier \#1: \\ Have we detected \\ rotation?};
    \draw [arrow] (pro1) -- (dec1);
    \node (o1) [no, below of=dec1, yshift=-2.5cm] {Non-detection};
    \draw [arrow] (dec1) -- (o1);
    \node (pro2) [process, align=center, right of=dec1, xshift=4cm] {Yes};
    \draw [arrow] (dec1) -- (pro2);
    \node (dec2) [decision, align=center, right of=pro2, xshift=4cm] {Classifier \#2: \\ Have we detected \\ an accurate period?};
    \draw [arrow] (pro2) -- (dec2);
    \node (o2) [no, below of=dec2, yshift=-2.5cm] {Non-detection};
    \draw [arrow] (dec2) -- (o2);
    \node (o3) [yes, right of=dec2, xshift=3cm] {Detection};
    \draw [arrow] (dec2) -- (o3);
\end{tikzpicture}
\caption{Flowchart describing the pipeline used in this work to detect rotation periods in single TESS sectors.}
\label{fig:flowchart}
\end{figure}

\begin{figure}
    \centering
    \includegraphics[width=0.8\textwidth]{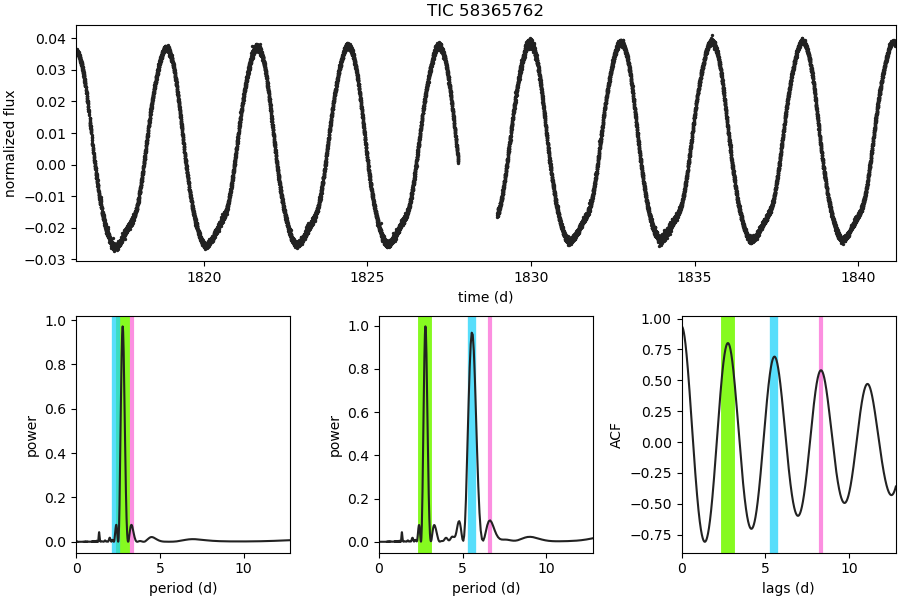}
    \caption{TESS single-sector light curve for TIC~58365762 (sector 19), exhibiting a clear signal of stellar rotation. The bottom panels show the LS periodogram, 2-term LS periodogram, and ACF of the light curve. Peaks A, B, and C are marked by vertical highlights of decreasing width, which are colored to guide the eye and make them distinct from the plotted functions.}
    \label{fig:dete}
\end{figure}

\begin{figure}
    \centering
    \includegraphics[width=0.8\textwidth]{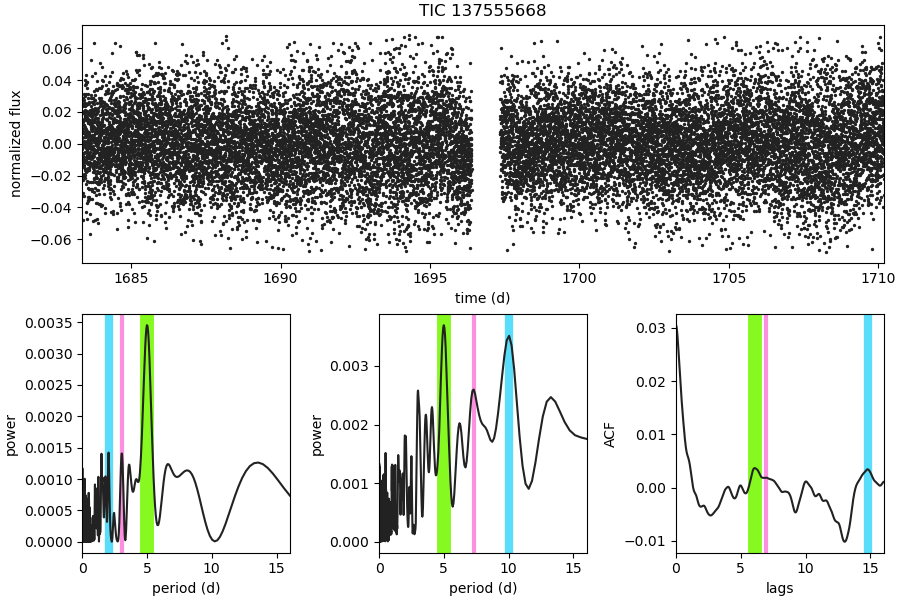}
    \caption{TESS single-sector light curve for TIC~137555668 (sector 14), dominated by noise, with no visually discernible signal of any kind. The bottom panels show the LS periodogram, 2-term LS periodogram, and ACF of the light curve. Peaks A, B, and C are marked by vertical highlights of decreasing width. Note that the LS peaks are very low amplitude and close to the noise level; also note the lack of periodic pattern in the ACF.}
    \label{fig:nond}
\end{figure}

\begin{figure}
    \centering
    \includegraphics[width=0.8\textwidth]{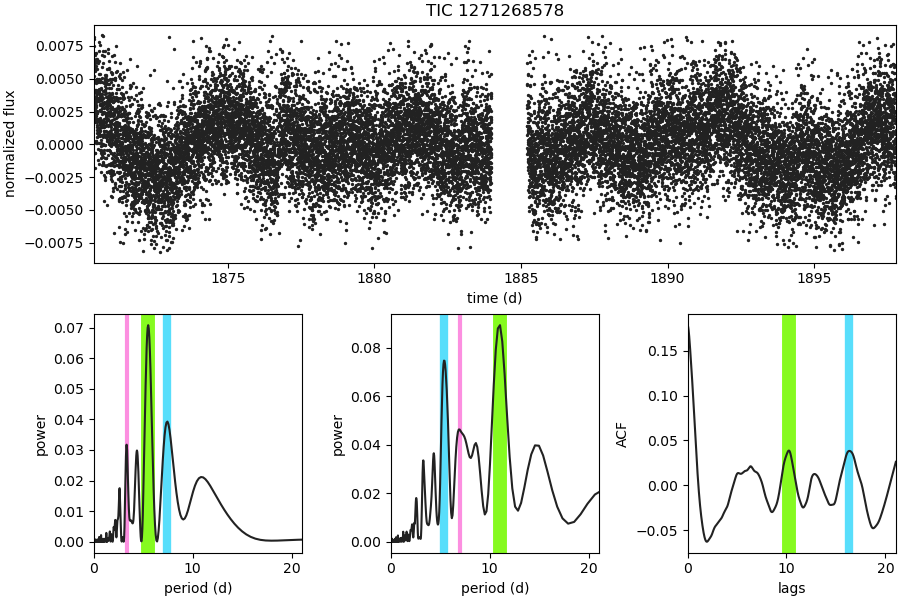}
    \caption{TESS single-sector light curve for TIC~1271268578 (sector 21), a marginal case with a period visible close to the noise level. The bottom panels show the LS periodogram, 2-term LS periodogram, and ACF of the light curve. Peaks A, B, and C are marked by vertical highlights of decreasing width. Note the lack of clarity in the ACF, and the 2-term LS detection of the double period. As will be discussed in Section~\ref{sec:results1}, this star is in fact a ``double-dipper.'' Due to the high noise level, the single-term LS does not detect the correct period in this case, though the sector does pass both classifiers as a detection of rotation.}
    \label{fig:marg}
\end{figure}

The parameters we use to classify targets are solely light curve features; that is, we do not use stellar parameters. This ensures that the random forests are not biased by the typical distribution of cool star rotation periods in magnitude and \teff\ space, and should be spread evenly across the stellar parameter space of the input catalog. We perform three main forms of analysis to determine these features, as introduced in Section~\ref{sec:intro}: two Lomb-Scargle (LS) periodograms as implemented in \textit{Astropy} \citep{collaboration_astropy:_2013, collaboration_astropy_2018} and an autocorrelation function (ACF) as implemented in \textit{Starspot} \citep{ruth_angus_2021_4613887}. Of the two LS periodograms, the first (Figure~\ref{fig:dete}, lower left panel) is the classic formulation, fitting single sine waves to points in a period-space grid to determine the amplitude at each period. We adopt the highest peak in this first LS periodogram as the star's rotation period --- initial testing (see also Section~\ref{sec:methods3}) showed this to be overall the most reliable measurement for the majority of targets, though we find it necessary to make adjustments in some cases, which will be discussed in Section~\ref{sec:results3}. The second LS periodogram is ``two-term'' (Figure~\ref{fig:dete}, lower middle panel), the main difference being that it fits the first two terms of a Fourier series (at each given frequency and its harmonic) per period in the grid, allowing for more robust detection of the primary period in multiperiodic systems. We detect peaks in the two LS periodograms and the ACF (Figure~\ref{fig:dete}, lower right panel) using a sorted peak-finding algorithm to avoid detecting sidelobes, and record the three highest-amplitude peaks in each case, termed A, B, and C. Figure~\ref{fig:dete} shows an example of peak detection for a visually clear rotation signal; for comparison, we show a TESS light curve dominated by noise and the measurements which contribute to its classification as a non-detection in Figure~\ref{fig:nond}, and a marginal case, harder to determine by eye, in Figure~\ref{fig:marg}.

From these detections, the features that we then include in our classifiers are the ratios between peaks A/B (Ratio 1), A/C (Ratio 2), and B/C (Ratio 3). We find that using ratios instead of periods allows the classifiers to work outside the range of periods represented in the training data, while still representing the characteristic harmonic patterns of rotational signals: the first three peaks in the power spectrum of a rotator are highly likely to be multiples of one another, whereas in a noise-dominated power spectrum their distribution across period space will be stochastic, and for eclipsing binaries, the ratios will be much smaller. We also include the amplitudes of each set of peaks A, B, and C. Note that despite what appears to be a coherent pattern by eye in Figure~\ref{fig:nond}, the non-detection, the amplitudes are extremely low. By contrast, the marginal case (Figure~\ref{fig:marg}) shows significantly higher amplitudes, especially compared to the characteristic noise level. Not all measurements will be possible for all stars; specifically, a noisy ACF may not exhibit any clear peaks, but this will aid its classification as a likely non-detection. In addition, our random forests include the root-mean squared error (RMS) and median absolute deviation (MAD) of each measured peak.

As well as period feature measurements, we include several other more general light curve features. From each LS, we measure the median level of power across the spectrum. A higher level of median power generally corresponds to higher background photon noise, which makes detection less likely. The median power also serves as a simple proxy for non-rotational stellar activity: if solar-like oscillations, eclipses, or transits are present in the periodogram, they will show up as a series of closely-clustered peaks, which raises the median power level. However, as stellar rotation presents in period-power space as a series of narrow, isolated peaks, the median power will be closer to the background noise level; that is, lower. One caveat for this is that the signals from compact or ellipsoidal binaries and classical pulsators appear similarly to stellar rotation in feature space. There will always be a degree of contamination risk from such objects if we only use the periodogram. However, in combination with other LS features, and the ACF, it is sufficient that the median power measure mainly serves to remove solar-like oscillators, eclipsing binaries, and transiting exoplanet hosts from the dataset.

Finally, we measured the combined differential photometric precision (CDPP) over a 2 hour timescale, using the CDPP implementation in Lightkurve \citep{collaboration_lightkurve:_2021}, which calculates scatter after the removal of long-term trends by using a low-pass filtered light curve. Noisier light curves will have higher CDPP, and we find that these targets are less likely to yield accurate period measurements in Classifier \#2. For comparison, we also experimented with including \rvar, the relative amplitude of variation, defined as difference in flux between the 95th and 5th percentiles. However, \rvar\ is dependent on the data product and light curve processing methods, and we found a significant difference in the distribution of \rvar\ between the \textit{eleanor} light curves from the FFIs and the TESS short-cadence light curves, such that it actively introduced noise into the classification process. CDPP, on the other hand, retained a constant distribution in both datasets. Nevertheless, we report \rvar\ as a useful parameter for understanding the data in our output catalogs (see Section~\ref{sec:results3}).

This totals 39 parameters in each random forest classifier. The random forest is still fast to generate and deploy despite this, and the use of more parameters allows for more information to go towards determining the feature space morphology of rotation in a single TESS sector.

\subsection{Comparison to other pipelines} \label{sec:methods2}

In this section we provide a brief overview of other software pipelines for detecting stellar rotation, to situate this work in the literature and highlight the unique features of our pipeline. There are multiple rotation pipelines both published and in development that extract periods from TESS data. However, the most complete catalogs we have to date are high-confidence detections among the Kepler corpus. The ROOSTER pipeline \citep{breton_rooster_2021} has been applied with success to Kepler data \citep{santos_surface_2021}, with a view to use it on TESS. Though this means we cannot yet compare the pipelines' efficacy directly, ROOSTER is noted here as the pipeline also uses random forest classifiers for vetting. ROOSTER vets in three stages: for detection, to remove compact binaries and classical pulsators, and finally to determine the correct period. ROOSTER uses 159 parameters, drawn primarily from an ACF and from a combination of the ACF and a wavelet transform, known as a composite spectrum. The pipeline has an accuracy of $>95\%$ for period detections; after incorporating a visual inspection stage, this increases to $>99\%$.

Turning to pipelines that have been applied to TESS data, we will first discuss the SpinSpotter pipeline \citep{holcomb_spinspotter_2022}, hereafter H22, which uses data from an ACF in a similar feature space formulation to our work. SpinSpotter fits a parabola to the peaks of the ACF to describe its morphology; several light curve parameters are also included. The pipeline then uses user-defined selection criteria to vet for the accuracy of a period detection. Unlike this work, SpinSpotter uses stitched TESS sectors to confirm single-sector detections, where available. The results from H22 are the most closely comparable to ours, which will be further detailed in Section~\ref{sec:results3}.

Work by \citet{kounkel_untangling_2022} (hereafter K22) searches for rotation periods for stars with known ages from the Theia associations \citep{kounkel_untangling_2020}, with the aim of improving gyrochronological relations. Like this work, the K22 pipeline operates on one TESS sector at a time, and uses a LS periodogram to search for evidence of rotation. They find that the detected rotation periods reinforce the spin-age relation as understood from initial Theia analysis, which underscores the utility of the LS periodogram for TESS analysis. However, both this study and H22 found that no method could reliably detect periods longer than $\sim$12 days, due to systematics introduced by the data downlink gap every sector.

Another major pipeline is the work by \citet{claytor_recovery_2022}, which uses a convolutional neural network to detect rotation periods in TESS. The use of synthetic data guarantees a 100\% accuracy rate among ground truth periods used to train the neural network, while allowing the authors to incorporate real TESS systematics and test on different versions of each light curve, such as with or without injected noise. They use the wavelet transform method to detect a period; the wavelet is then processed by the neural network to vet the detection. They are able to recover 63\% of periods $<$50~d with up to 10\% accuracy. This pipeline was recently applied to actual TESS light curves in the Southern Continuous Viewing Zone \citep{claytor_tess_2023}, retrieving 7,971 rotation periods up to a length of 80~d. As our study uses shorter time series and aims to recover shorter periods, it is hard to compare our results directly. However, these significant results provide a complementary data product to ours.

A recent pipeline released by \citet{fetherolf_variability_2022} is designed not for rotation but to detect stellar variability in general in TESS. It merits mention here as, by coincidence, the authors also use the two LS periodograms and an ACF as described above. They then classify variability based on peak significance. They analyze the first 26 sectors of TESS and detect 89,448 variable targets. As this catalog contains similar data processed in a similar way to our work but with a different focus, we foresee that this work could be useful in conjunction with rotation catalogs for comparative studies.

\subsection{Training data} \label{sec:methods3}

We trained our classifiers on the rotation period catalog plublished by \citet{oelkers_variability_2018} (O18), which comprises of detections from a ground-based survey of TESS targets, using the KELT telescope \citep{pepper_kilodegree_2007}. This includes a total of 64,039 rotators and 58,598 non-detections. \citet{oelkers_variability_2018} identified periods using a LS periodogram, and used light curve metrics such as peak significance and light curve dispersion to vet for accuracy. We removed known eclipsing binaries (EBs) from the training set by crossmatching with stellar designations on Simbad \citep{wenger_simbad_2000}. We produced light curves for all targets in the O18 catalog from the TESS full frame images (FFIs) using \textit{eleanor} \citep{feinstein_eleanor_2019}, discarding any targets which could not be processed due to their proximity to the edges of the FFIs, resulting in a final sample of 57,480 rotators and 53,264 non-detections. Using FFI light curves for training ensures limited overlap with the data in our final catalog (see Section~\ref{sec:methods4}).

\begin{figure}
    \centering
    \includegraphics[width=\textwidth]{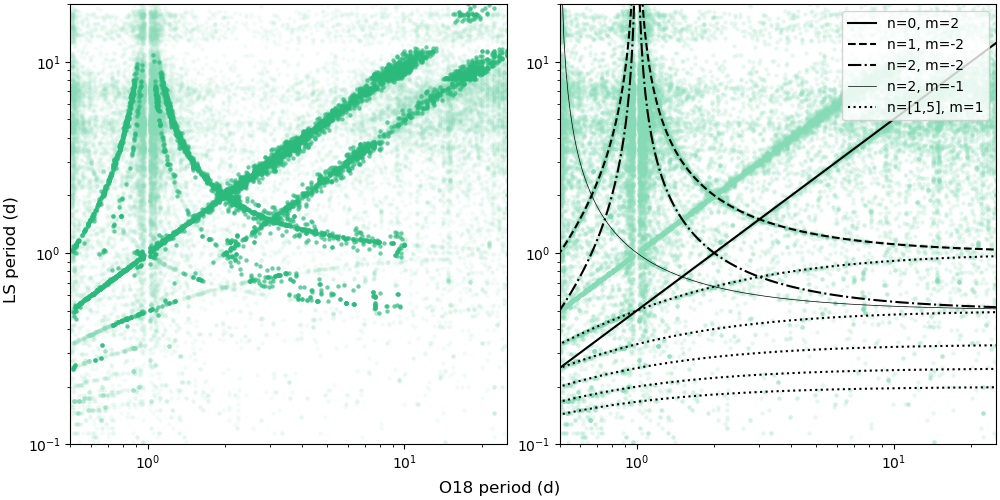}
    \caption{Left: period accuracy in the training set for Classifier \#2, with O18 periods plotted on the x-axis and detections from this work on the y-axis. Periods classified as accurate for training purposes are shown in the darker color. Right: failure modes used to select accurate periods in the training set, with the unused failure modes also highlighted. We expect that ground truth periods lying along these lines are likely aliases, due to the systematic feature at 1~day from Earth's rotation.}
    \label{fig:failmodes}
\end{figure}

After extracting parameters from the \textit{eleanor} light curves, we classified O18 detections  below 2~d with an LS amplitude of $<$0.5 (dimensionless, with values between 0 and 1) as non-detections for training both Classifier \#1 and \#2. This removed a large amount of noise in the O18 catalog, without biasing our classifiers against periods below 2~d, as we will show. We divided up the O18 data using a standard 70\% training/20\% validation/10\% testing breakdown. To train Classifier \#2, we required an LS period detection within 15\% of the O18 value. We also allowed periods within 15\% of three LS failure modes, a common effect in ground-based data where an alias of the Earth's rotation period is detected as the dominant peak due to window function effects. We defined failure modes based on the equation given in \citet{vanderplas_understanding_2018} for the case where there is a strong contaminating signal at 1~day:

\begin{equation}
    P_{LS} = \bigg| \frac{m}{P_{O18}} + n \bigg| ^{-1}
\end{equation}

\noindent with modes at n=0/m=2 (the harmonic/half-period), n=1/m=-1, and n=2/m=-2, selected empirically to match failure patterns present in the training data, as shown in Figure~\ref{fig:failmodes}. There are also periods which lie along the n=2/m=-1 failure mode, and a series of failure modes at n=[1,5]/m=1. We found that using these additional periods in Classifier \#2 yielded a high true negative rate, but extremely low true positive rate, and therefore we excluded them from the final formulation.

\begin{figure}
    \centering
    \includegraphics[width=0.8\textwidth]{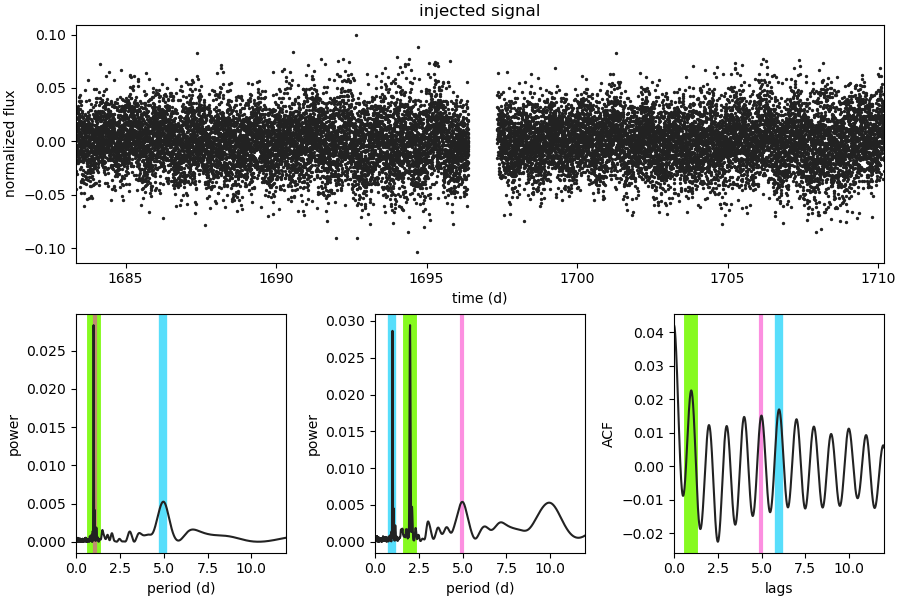}
    \caption{Low-amplitude sinusoids of 1 day and 5 days injected into the TESS single-sector light curve for TIC~137555668 (sector 14). The bottom panels show the LS periodogram, 2-term LS periodogram, and ACF of the light curve. Peaks A, B, and C are marked by vertical highlights of decreasing width. Note particularly that, in a marginal case such as this, the LS periodogram still accurately detects the predominant signal of 1 day.}
    \label{fig:inj}
\end{figure}

It cannot be taken for granted that failure modes address all possible cases of aliasing and incorrect period detection in a Lomb-Scargle formulation \citep[e.g.][]{dawson_radial_2010}. This has recently been noted in the search for spot-induced modulation in ground-based time series data from the Zwicky Transient Facility \citep{getman_pre-main_2023}, which suggests particular caution using a ground-based dataset as our training set. To a large degree, this is avoided in our pipeline by re-measuring ground truth periods in the O18 data, assuming that the ground truth is a detection of rotation and not an alias. To confirm our choice of the primary peak in the LS periodogram as the new ground truth period, we performed several tests using sinusoidal signals injected into the noise-dominated TESS light curve shown in Figure~\ref{fig:nond}. We show an example of this process in Figure~\ref{fig:inj}, where we have injected a 1~day sinusoid with at an arbitrary amplitude of 5\% of the total flux, and a 5~day period with an amplitude of 0.5\% of the total flux. Notably, the LS periodogram, which we adopt as our ground truth period in both the training data and the pipeline, correctly returns the dominant period. Our tests of marginal cases confirm that the single-term LS periodogram in particular performs as expected for detecting sinusoidal variation even in cases where the noise profile dominates. In Section~\ref{sec:results1} we will elaborate on the utility of the 2-term LS and the ACF in determining the validity of LS results for more complex cases.

\begin{figure}
    \centering
    \includegraphics[width=\textwidth]{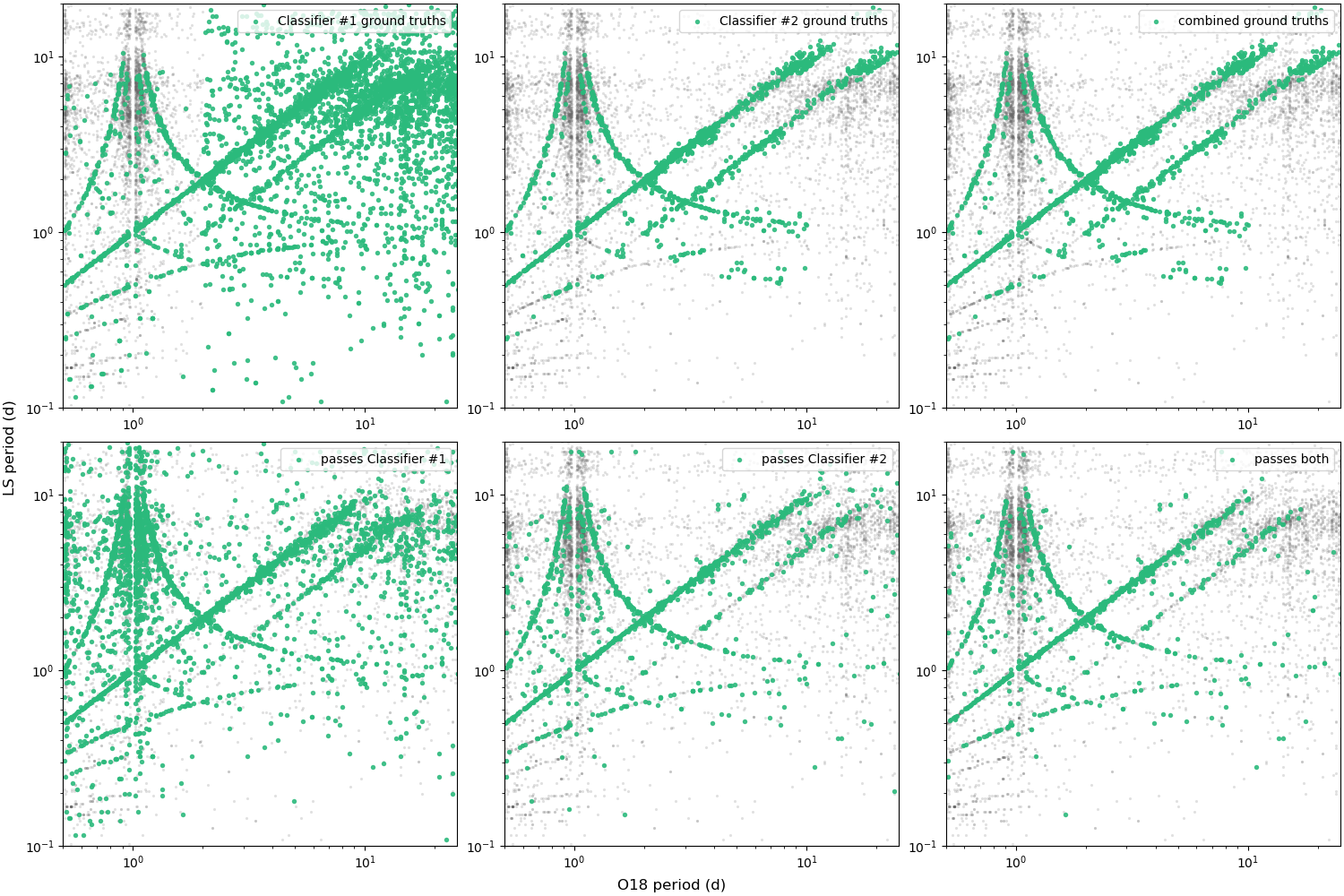}
    \caption{Top row: ground truths for the validation dataset for Classifier \#1 (left), \#2 (center), and both combined (right). Bottom row: the results of applying Classifiers \#1 (left) and \#2 (center) to the validation dataset, and targets that pass both classifiers (right). Periods that pass classification are shown in green, larger points.}
    \label{fig:validated}
\end{figure}

We tuned the training set based on results from the validation dataset, 20\% of the O18 catalog, by classifying validation data independently using the same metrics as above. We optimized the number of estimators used in each random forest, choosing three as a number that maximizes the percentage of true positives and true negatives for both classifiers. This is an atypically low number of estimators to use in such a problem --- however, we find a greater benefit in an approach that is more permissive, allowing for a high-accuracy output catalog that is largely free from contamination. Later, we will discuss the process of making an additional amplitude cut for output periods, as the classifier is of course not 100\% accurate.

We show validation results for each classifier, and for targets that passed both, in Figure~\ref{fig:validated}. In the validation dataset, Classifier \#1 correctly identified 41\% of O18 rotators (the true positive rate) and 83\% of non-detections (true negatives). Classifier \#2 yielded 56\% of periods marked as accurate and had a true negative rate of 95\%. The high true negative rates for both classifiers gives us confidence in the pipeline's ability to disregard both non-rotators and inaccurate period measurements. It is important to note that these true negative classifications included ground truth non-detections, though Classifier \#2 was only trained on ground truth detections (many of which are discarded for lying off the one-to-one line and failure modes, and thus classified as inaccurate periods for training purposes) as we cannot assess the accuracy of samples with no rotation period.

Crucially, of the targets in the validation sample classified as both rotators and accurate periods, 70\% of ground truth double-positives (894 of 1,276) pass both classifiers. This is visualized in the two right-hand panels of Figure~\ref{fig:validated}. We also calculated true positive rates for ground truth periods less than 10d: Classifier \#1 yielded 58\% of rotators, and Classifier \#2 yielded 60\% of accurate periods. Combined, they yielded 74\% of double-positive ground truths. We note, additionally, that short ($<$~10~d) periods, such as those we aim to find in individual TESS sectors, constituted a small percentage of the training data: 15\% of periods under 10~d were considered true rotation detections, and the same proportion of the sample was considered to have an accurate period. We will show later that, due to the size and robustness of the training data and the edsign of our feature space, this did not bias our final classification process against short periods.

As mentioned previously, we reclassified 1,129 EBs with rotation periods in the training set as non-rotators for Classifier \#1 and inaccurate periods for Classifier \#2. We compared the results of each classifier before and after reclassification, based on 323 EBs identified in the validation set. We found an improvement from 33 to 24 EBs passing Classifier \#1 once EBs in the training set were reclassified, and 43 to 32 passing Classifier \#2. Ultimately, only 3 of 323 EBs passed both classifiers, which suggests that the pipeline is robustly able to remove EBs based on their distinct feature space morphology.

Finally, after the completion of all feature engineering, we ran the pipeline on the testing set, 10\% of the overall training sample, for true blind validation. We found that Classifier \#1 correctly identified 41\% of rotators and had an 82\% true negative yield; Classifier \#2 yielded 55\% true positives and 95\% true negatives. Both classifiers combined yielded 72\% double-positive ground truths. Finally, we note that of 160 EBs identified in the testing set, only 7 passed both classifiers. These results are almost precisely comparable to the results from the validation set, which confirms that our vetting algorithm has been robustly constructed and optimized to detect short-period signals of rotation in 27~d TESS sectors.

\begin{table}
\centering
\begin{tabular}{rlrl}
\textbf{Classifier \#1 Feature} & \textbf{\% Importance} & \textbf{Classifier \#2 Feature} & \textbf{\%Importance} \\
LS amplitude A	&	5.90 &	LS amplitude A	&	25.56 \\
LS2 amplitude B	&	5.77 &	LS2 amplitude B	&	14.60 \\
LS2 amplitude A	&	4.45 &	LS2 amplitude C	&	3.80 \\
LS median power	&	3.62 &	LS amplitude C	&	2.42 \\
LS2 median power	&	3.62 &	LS2 median power	&	2.32 \\
LS2 amplitude C	&	3.29 &	LS2 amplitude A	&	2.27 \\
CDPP	&	3.27 &	ACF amplitude A	&	2.18 \\
LS amplitude C	&	3.09 &	ACF amplitude B	&	2.17 \\
LS RMS B	&	3.09 &	LS ratio 2	&	2.13 \\
ACF RMS B	&	3.00 &	LS ratio 1	&	2.04 \\
ACF amplitude A	&	2.92 &	LS2 ratio 1	&	1.97 \\
LS2 ratio 1	&	2.77 &	LS median power	&	1.85 \\
LS2 MAD B	&	2.69 &	LS2 ratio 2	&	1.84 \\
LS2 ratio 3	&	2.68 &	LS2 ratio 3	&	1.83 \\
LS ratio 2	&	2.65 &	ACF MAD C	&	1.83 \\
LS2 ratio 2	&	2.60 &	LS amplitude B	&	1.82 \\
LS amplitude B	&	2.56 &	LS2 RMS B	&	1.61 \\
LS ratio 1	&	2.43 &	CDPP	&	1.60 \\
LS ratio 3	&	2.42 &	LS ratio 3	&	1.58 \\
LS MAD A	&	2.39 &	LS2 RMS A	&	1.54 \\
ACF amplitude B	&	2.24 &	ACF RMS A	&	1.54 \\
LS2 RMS B	&	2.15 &	ACF ratio 1	&	1.50 \\
LS2 RMS A	&	2.12 &	LS2 MAD C	&	1.43 \\
ACF RMS A	&	2.10 &	ACF RMS B	&	1.39 \\
ACF MAD A	&	2.08 &	LS MAD B	&	1.38 \\
LS2 MAD A	&	1.96 &	ACF MAD A	&	1.33 \\
LS RMS A	&	1.92 &	LS MAD A	&	1.32 \\
LS MAD B	&	1.92 &	LS RMS A	&	1.31 \\
LS2 MAD C	&	1.91 &	LS2 MAD B	&	1.26 \\
ACF MAD B	&	1.91 &	ACF RMS B	&	1.25 \\
ACF RMS B	&	1.89 &	LS MAD C	&	1.23 \\
LS2 RMS B	&	1.88 &	LS2 RMS B	&	1.21 \\
ACF MAD C	&	1.88 &	ACF MAD B	&	1.21 \\
LS MAD C	&	1.87 &	LS RMS B	&	1.13 \\
ACF ratio 1	&	1.81 &	LS2 MAD A	&	1.13 \\
LS RMS B	&	1.77 &	LS RMS B	&	1.12 \\
ACF amplitude C	&	1.37 &	ACF amplitude C	&	0.90 \\
ACF ratio 3	&	1.06 &	ACF ratio 2	&	0.73 \\
ACF ratio 2	&	0.98 &	ACF ratio 3	&	0.69    
\end{tabular}
\caption{Percentage importance of all 39 features ranked for both classifiers.}
\label{tab:importances}
\end{table}

\begin{figure}
    \centering
    \includegraphics[width=0.8\textwidth]{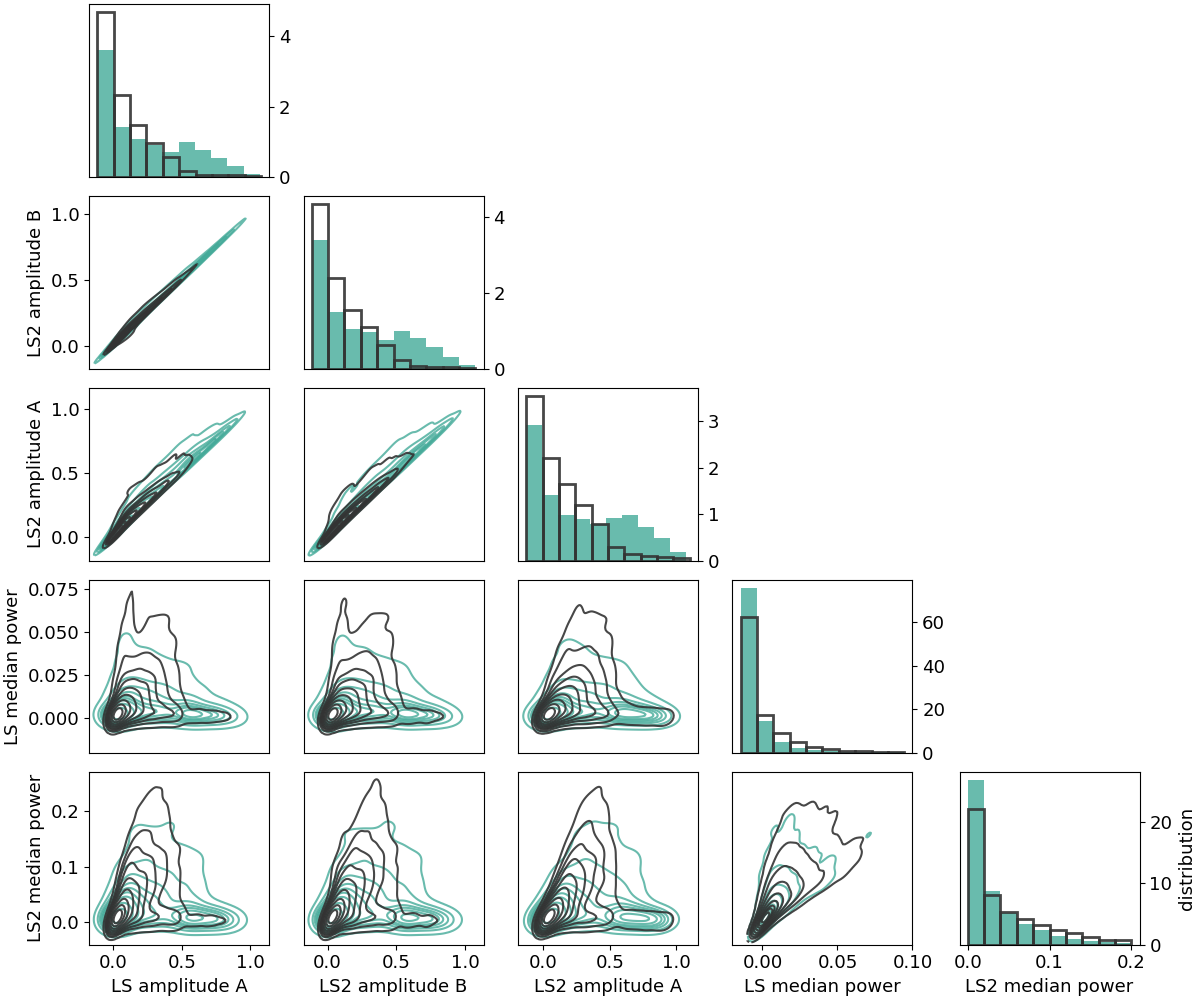}
    \caption{Corner plot showing the interactions between the five most important features in Classifier \#1, including density histograms for each feature. Measurements that passed the classifier, i.e. detections of rotation, are shown in teal, and non-detections are shown in black.}
    \label{fig:c1c}
\end{figure}

\begin{figure}
    \centering
    \includegraphics[width=0.8\textwidth]{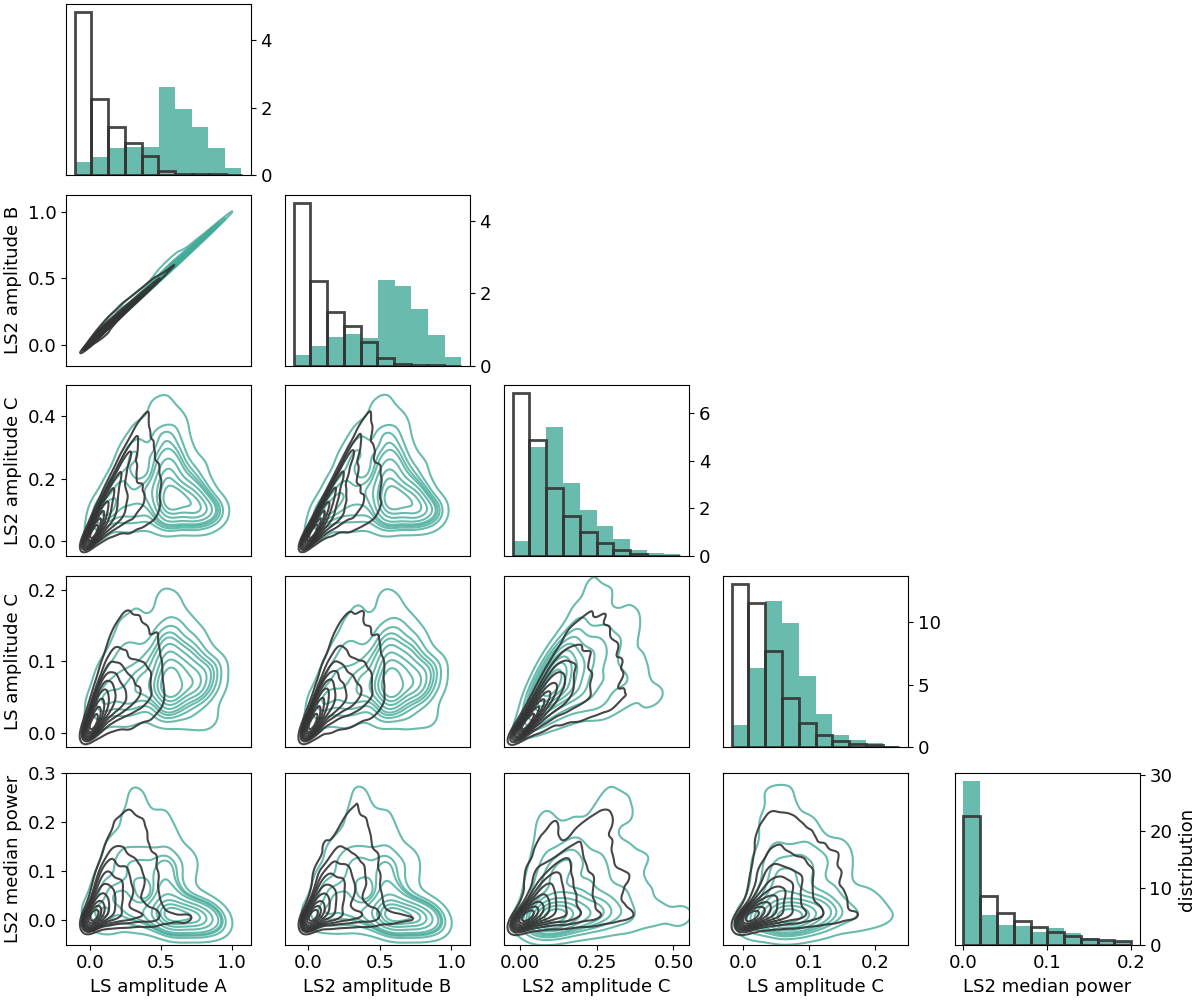}
    \caption{Corner plot showing the interactions between the five most important features in Classifier \#2, including density histograms for each feature. Measurements that pass the classifier, i.e. accurate periods, are shown in teal, and periods deemed inaccurate are shown in black.}
    \label{fig:c2c}
\end{figure}

The implementation of random forests in \textit{scikit-learn} provides a function to retrieve the relative importance of each feature in a classifier. In Table~\ref{tab:importances}, we show all 39 features ranked with relative importances for both classifiers. To visualize this, Figures~\ref{fig:c1c} and~\ref{fig:c2c} show corner plots of the validation data for the five most important features in each classifier and how they correlate with one another. Of note, the most important feature for both classifiers is the amplitude of peak A in the one-term LS periodogram. This contributes to our choice of LS peak A as the target's rotation period, as amplitude is correlated with the significance of the detection. Indeed, the amplitudes for all peaks in both LS periodograms are ranked highly for both classifiers; though this is not to discount the ACF, which is across the board more important in its amplitudes than corresponding peak ratios. For Classifier \#1, the noise proxy features (CDPP and LS median power) are ranked more highly than they are in Classifier \#2. This can be understood in terms of the relationship between median power and diffuseness of signal, e.g. distinguishing between noise and rotation, and the fact that a lower CDPP reflects a more coherent light curve. However, Classifier \#2 largely preferences amplitudes; in fact, the amplitude of the target's rotation period is weighted at 26\% importance, meaning that a large quantity of ``decisions'' in the random forest trees depend on a high-amplitude detection. In the training set, the drop-off in detections appears around an amplitude of 0.5 (dimensionless). This is the arbitrary cut-off amplitude used at several points in the training set, as explained above, but of course the classifiers do not know this --- rather, the re-emergence of that number in the validation data confirms that random forest algorithms are well-suited for this classification problem and that our classifiers have a high fidelity to the training data.

\subsection{Input Catalog} \label{sec:methods4}

In this study, we analyzed TESS 2 minute cadence light curves from the first 26 sectors of TESS \textbf{\citep{tess_doi}}. Using freely available, pre-processed data reduced computational time and streamlined target selection for this study. We performed a 3-sigma outlier clip on all light curves, to remove outliers due to data quality issues, and eclipses. We then normalized the light curves to be centered at zero --- this ensured that the ACF functioned as expected.

To limit the number of input targets to only those most likely to yield detections, we imposed cuts based on stellar parameters in the TESS Input Catalog (TIC). This also resulted in a small reduction in computational time, but it was insignificant compared to the overall size of the short cadence dataset. We first removed targets with a TIC effective temperature greater than 7,000K. (We included targets with no available \teff.) We chose 7,000K to be consistent with the upper limit of the results in \citet{santos_surface_2021}. At the higher end of this \teff\ range, our sample is susceptible to contamination by compact or ellipsoidal binaries and classical pulsators. Whereas ROOSTER \citep{breton_rooster_2021} uses a third random forest to remove these, detections from our pipeline run a small but extant risk of confusion, as the light curve morphologies are so similar in both cases. We count on the majority of these cases to be removed by Classifier \#1, as described in Section~\ref{sec:methods1}. To remove giants from the catalog, we imposed a limit of 0th Gaia absolute magnitude, calculated from Gaia DR2 \textit{G}-band magnitude and parallax \citep{gaia_hr_2018}. Finally, we remove extended sources included in the TIC from our catalog.

Additionally, we wanted to limit our targets to those with photometry conducive to good light curve feature detection, so we avoided saturated targets by cutting TIC entries with a TESS magnitude brighter than 5th. This is a close match to the saturation ``limit'' naturally imposed by the O18 training set, 4th TESS magnitude. The training set does not include stars fainter than 12th TESS magnitude; we cut at 16th magnitude to allow for low-magnitude serendipity, while also avoiding the possibility of targets so faint that all the signal in their light curves is due to contamination. Fewer than 5\% of targets in each sector are below 16th magnitude. Thus, from a possible 523,964 sectors, our input catalog contained 432,704 sectors for analysis, which covers 194,336 distinct targets.

\section{Results \& Discussion} \label{sec:results}

\subsection{Detecting rotation} \label{sec:results1}

We detected stellar rotation --- that is, a detection which passed both stages of vetting --- in 17,110 of 456,356 light curves in the input catalog. This was the result of 123,358 measurements passing Classifier \#1, and 34,579 passing Classifier \#2. This corresponds to a yield of 3.75\% per sector and 5.82\% per all targets surveyed. Noting the importance of the LS periodogram's peak amplitude in classification, we made a further cut at an amplitude of 0.01 (dimensionless) to remove 310 likely false positives, bringing us to a final catalog of 16,800 detections over 11,010 targets. While this seems, on the surface, to be a low figure, the results are in line with our expectations based on prior studies and the nature of individual TESS sectors. The overwhelming majority of periods detected are under 12 days --- we will discuss the small number of outliers later in this section --- as detection of longer periods is unlikely in TESS data, even in stitched light curves containing multiple sectors. It is likely that a significant fraction of the stars not captured by our pipeline simply do not have such short rotation periods. For stars that have consecutive observations, such as those in the continuous viewing zones, it may be possible to confirm detections of longer periods. However, because our pipeline was designed to analyze individual sectors, this essentially imposed a detection limit. Finally, our random forest classifiers were designed stringently to avoid false positives: thus, despite the low yield, we can be confident in the fidelity of our detections.

\begin{figure}
    \centering
    \includegraphics[width=\textwidth]{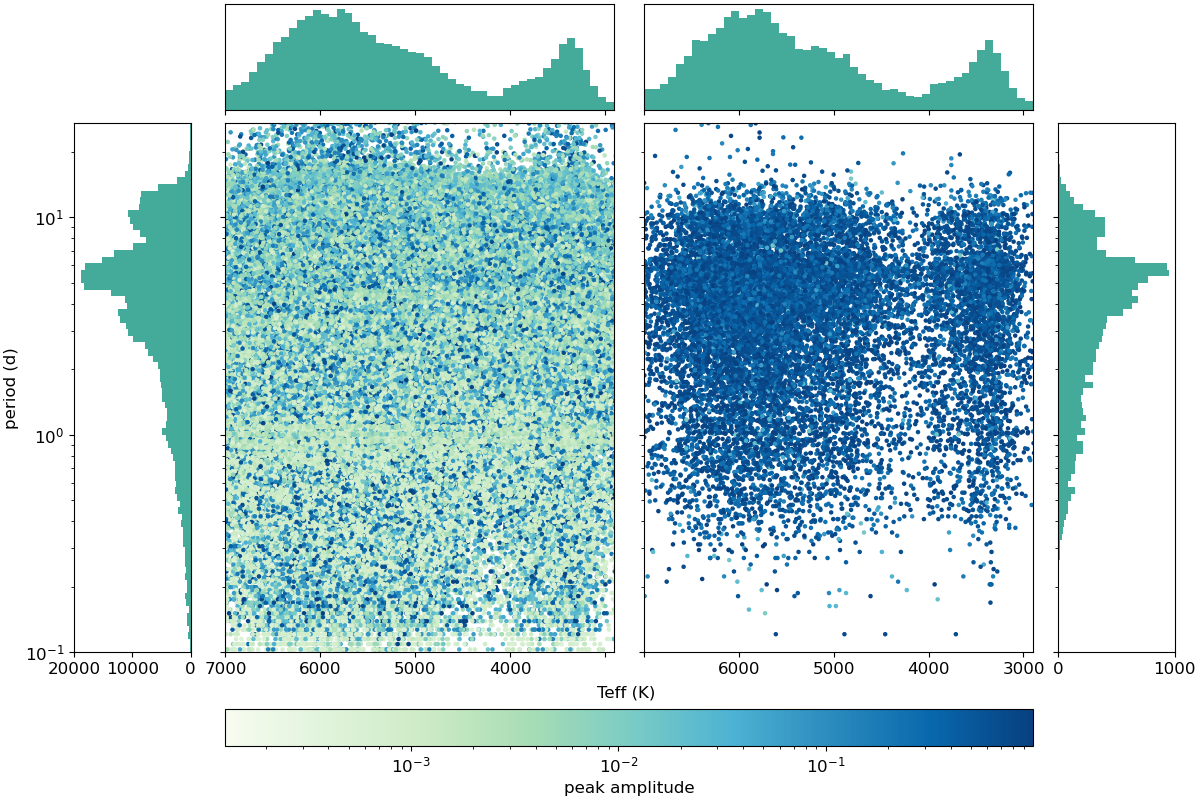}
    \caption{Periods measured in single sector light curves plotted against their effective temperatures. Periods that passed neither classifier are shown on the left, and those that passed both (i.e. detections) are shown on the right. The color scale indicates the amplitude associated with each detection. Note that 11,907 points are not plotted, representing 11,107 targets which do not have a value of \teff\ listed in either the TIC or Gaia DR3. This amounts to 2.8\% of the sample. Histograms at either side show the distribution of periods, and across the top we show the \teff\ distribution, which is reassuringly uniform for detections and non-detections.}
    \label{fig:pteff}
\end{figure}

In Figure~\ref{fig:pteff}, we show the per-sector periods detected by our pipeline plotted against \teff, where available in the TIC or Gaia DR3 \citep{gaia_gaia_2023}, color-coded by the amplitude of the period (or, in the case of non-detections, simply the highest amplitude present in the LS.) It is worth noting that, although naturally detection amplitudes skew larger, there is no hard trace of the 0.5 (dimensionless) amplitude cut imposed in various regions of feature space while training the random forest. We also note that, as expected (Section~\ref{sec:methods1}), detections are evenly distributed across \teff, reinforcing the lack of temperature dependence in our classification pipeline. This distribution also reflects the systematic underdensity of K dwarfs in the TIC, also seen in \citet{holcomb_spinspotter_2022}. We can see the fast-rotating M dwarf sequence in period-\teff\ space, and were able to detect a small amount of very fast rotators. We cannot observe the intermediate period gap in this sample, as it occurs at higher periods than we are sensitive to. Figure~\ref{fig:pteff} shows each detection as an individual point, meaning that targets with measurements across multiple sectors that pass both classifiers are represented multiple times. This has the effect of ``blurring'' the distribution, as measurements are not necessarily the same across sectors. This effect becomes less evident in our catalog when consolidated across targets, which we will discuss in Section~\ref{sec:results3}.

\begin{figure}
    \centering
    \includegraphics[width=0.7\textwidth]{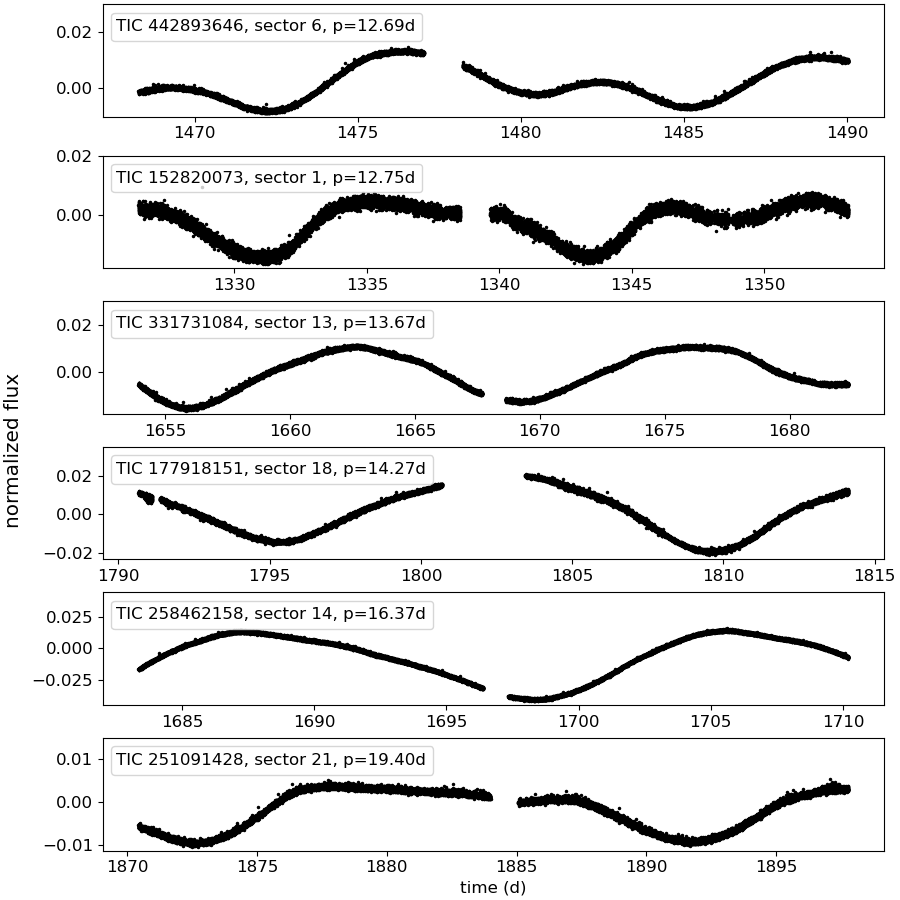}
    \caption{A selection of long ($>$12~day) periods which passed both classifiers, and were clear enough to visually confirm.}
    \label{fig:long}
\end{figure}

Among both detections and non-detections, our pipeline returned few periods above 12 days, upholding the theoretical detection limit. Detections above 12~d may be doubles of the true period (subharmonics) or aliases, or simply an inaccurate period measurement that has erroneously passed the second classifier; however, as the LS periodogram is essentially a pattern-matching algorithm, it can return a ``correct'' period measurement for a high signal-to-noise rotator based on a well-fitting partial match to a sinusoid. We observed this phenomenon in our training data. As there are only 309 targets with periods reported over 12~d, we performed visual inspection to determine that 30 targets appear to be genuine and accurate detections of longer-period rotators. We show a selection of these in Figure~\ref{fig:long}. We discarded 101 targets which erroneously passed one or both classifiers. The remaining $>$12~d detections are from multi-sector targets in the input catalog, and have $<$12~d detections in other sectors; visual inspection reveals that these spurious results tended to be subharmonics of the correct period, which was represented in another sector. In these cases, we discard the sectors with detections over 12~d in our consolidated catalog, leaving us with a final catalog of 10,909 targets.


\subsection{Output Catalog} \label{sec:results3}

\begin{table}
\centering
\begin{tabular}{cc}
\textbf{Available sectors} & \textbf{Incidence} \\
\textbf{per target} & \textbf{in C24} \\
1 & 8,364        \\
2 & 1,424        \\
3 & 498            \\
4 & 266          \\
5 & 160         \\
6 & 108               \\
7 & 63  \\
8 & 62        \\
9 & 32           \\
10  & 16          \\
11  & 8         \\
12  & 8          \\
13  & 1           
\end{tabular}
\caption{The C24 catalog includes median periods where there are multiple measurements per TESS sector. Here, we show the number of targets per number of sector measurements.}
\label{tab:dist}
\end{table}

We present a consolidated catalog (abbreviated in this work as C24) of 10,909 targets with detected rotation periods, available as supplementary material to the online edition of this paper. For targets with multiple sector detections, we report a median period. We find that a median better captures the distribution of periods in multi-sector detections than a mean: when the detected period differs significantly across sectors, the larger measurements are almost always double the smaller. The C24 catalog contains the median period, an adjusted period for 1,796 stars flagged as half-periods, TESS magnitude, and \teff\ (where available) for each target, as well as the number of distinct sector detections per target, and a standard error of the mean (SEM) where applicable. We also include a flag and adjusted periods for potential half-period (harmonic) detections, automatically applied to all detections where the 2-term LS and the ACF both return a period that is twice the LS period, which we will discuss in further detail later in this section. We break down the number of sectors with detections of rotation per target in Table~\ref{tab:dist}. It is worth noting that 8,364 targets are single-sector measurements; in general, where multiple sectors of data are available, one or more sector measurements failed one or both classifiers. A full catalog, including all 39 light curve parameters measured for each individual sector, is available on Zenodo: \href{https://zenodo.org/records/10684613}{DOI:10.5281/zenodo.10684613}.

\begin{figure}
    \centering
    \includegraphics[width=\textwidth]{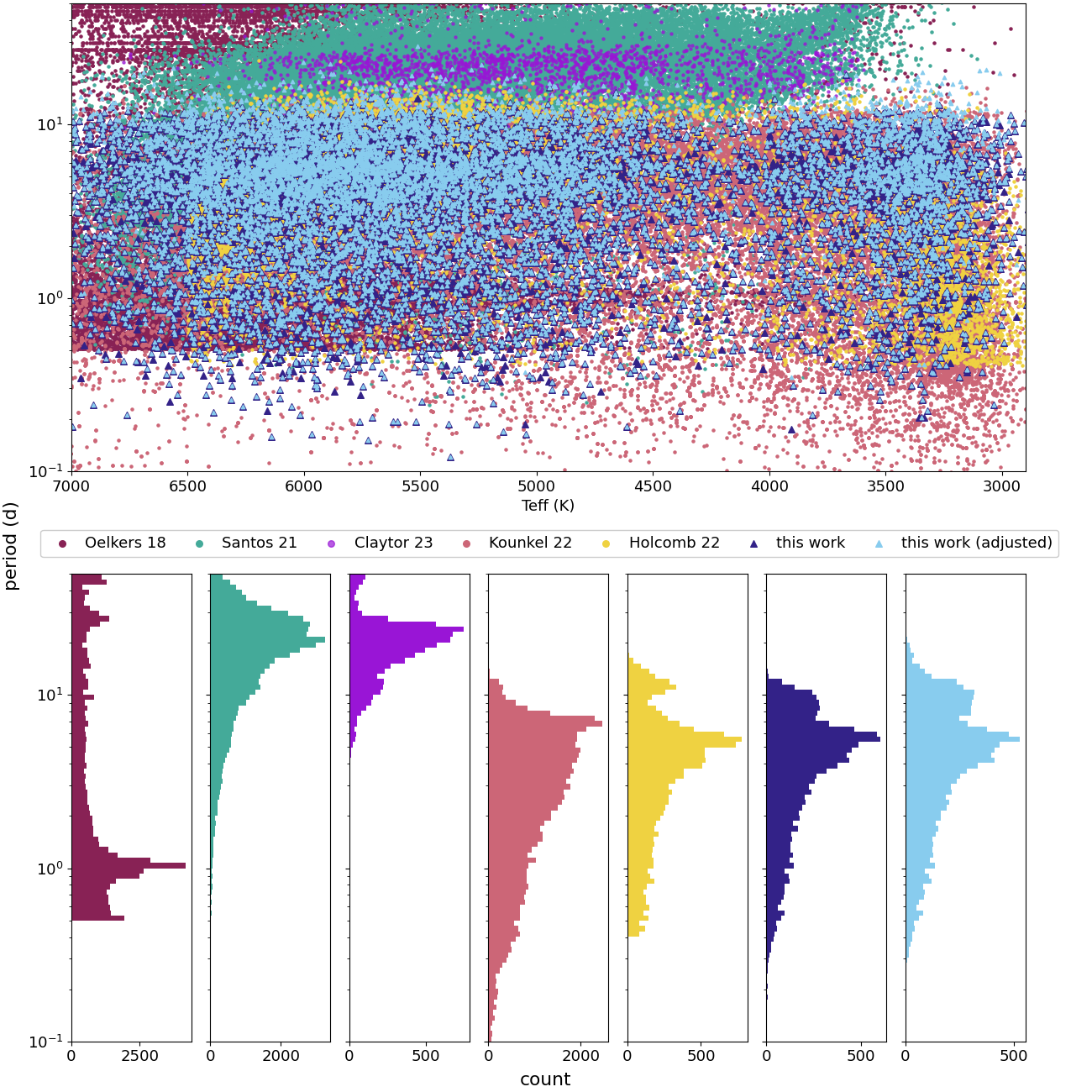}
    \caption{Comparison between this work (LS periods in dark blue and adjusted periods in light blue), \citet{oelkers_variability_2018} (burgundy), \citet{santos_surface_2021} (teal), \citet{claytor_tess_2023} (purple), \citet{kounkel_untangling_2022} (pink), and \citet{holcomb_spinspotter_2022} (yellow).}
    \label{fig:catalogs}
\end{figure}

We compare the C24 catalog to several other large catalogs of rotation periods for TESS and Kepler in Figure~\ref{fig:catalogs}. Our catalog probes overall lower temperatures and periods than the O18 catalog, our training set, which covers \teff\ up to 14,000~K and periods up to 50~d. As noted in Section~\ref{sec:methods3}, this has no impact on our ability to detect rotation at lower temperatures and periods, because light curve structure remains broadly the same for spot-induced rotation signals at different periods. In particular, we can compare our results to the two closest comparisons, \citet{kounkel_untangling_2022} (K22) and \citet{holcomb_spinspotter_2022} (H22). Due to the limitations of working with TESS data, both catalogs echo our findings: there is a sharp drop-off in detection over 10 days. Whereas K22 used full-frame image light curves, H22 analyzed the same corpus of data as our work --- short-cadence light curves from the first 26 TESS sectors, with the addition of stitched light curves where available, finding 13,504 rotation periods that passed their chosen selection criteria, and reporting that 89\% were 9 days or less. For comparison, 91\% of periods in the full catalog and 89\% of stars in the consolidated catalog are $<$9~d. The similarity of these figures confirms that there is a common detection limit when working with TESS data, independent of method and light curve length.

Another point of comparison with H22 is the systematic over-detection of periods around 6 days, as can be seen in Figure~\ref{fig:catalogs}, present despite our pipelines using different detection methodologies. We visually inspected a selection of periods between 5 and 7 days, and found that the only apparent false positives were low amplitude ($<$0.01) detections, which we unilaterally removed from the catalog as previously discussed. It is possible that some of these could be half-periods for 12--13 day periods --- these are less likely to be detected or pass vetting, as they are at similar frequencies as the data downlink gap. It may be that we can easily see half-periods where there is complex structure present at a primary period of 12--13 days. This is upheld to an extent by the diminution of the $\sim$6 day systematic significantly in the adjusted periods, and an increase in periods over 11 days.

\begin{figure}
    \centering
    \includegraphics[width=\textwidth]{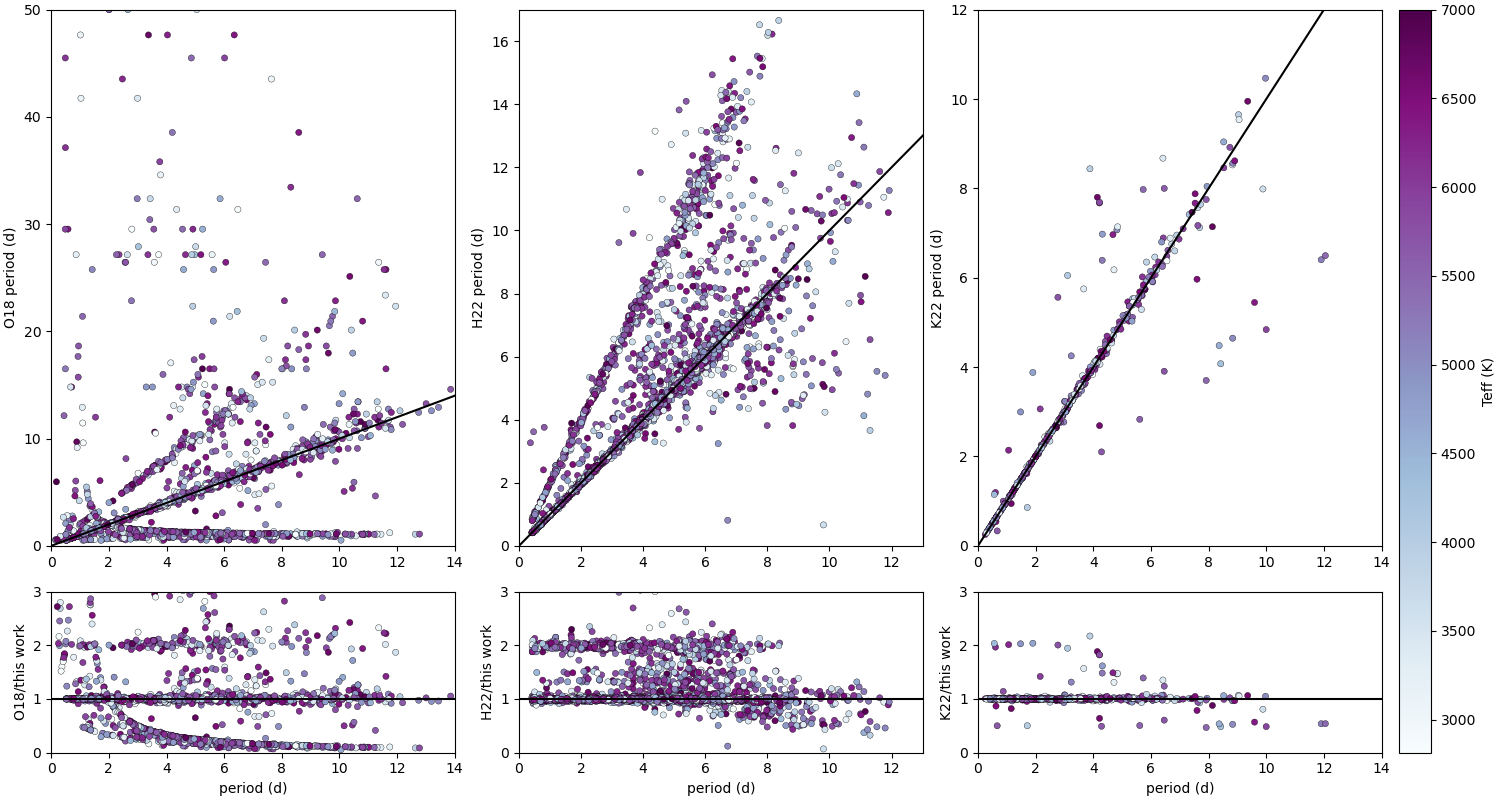}
    \caption{TESS rotation periods from our consolidated catalog compared with \citet{oelkers_variability_2018} (left), \citet{holcomb_spinspotter_2022} (center),  and \citet{kounkel_untangling_2022} (right), with residuals shown in the bottom plots. Axes are truncated to exclude a small number of outliers, and points are color-coded by the number of distinct sectors used to calculate each median period.}
    \label{fig:xmatchres}
\end{figure}

Crossmatching C24 with these other catalogs, we found that 1,884 of our targets are shared with O18, 4,864 with H22, and 468 with K22. C24 contains new rotation periods for 4,936 targets. This includes 250 rotation periods below 0.5~d, a region which has only been previously probed by K22. In Figure~\ref{fig:xmatchres}, we show periods where a measurement is present in both C24 and O18, H22, and K22 respectively. We find very good correspondence with the K22 catalog --- to be expected, given that K22 uses the LS periodogram to detect periods --- and we observe the same LS failure modes in the O18 overlap as in the training set.

\begin{figure}
    \centering
    \includegraphics[width=0.9\textwidth]{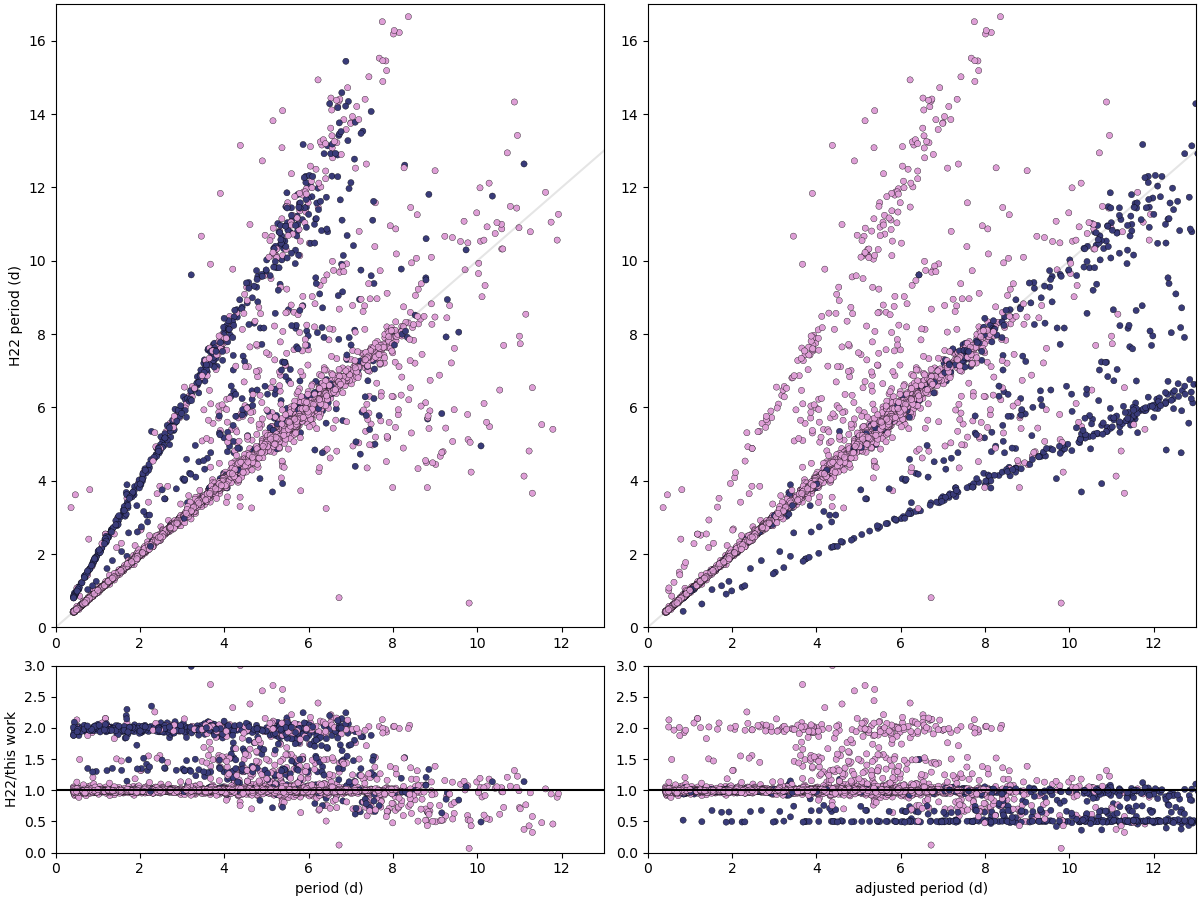}
    \caption{TESS rotation periods from our consolidated catalog compared with \citet{holcomb_spinspotter_2022} before (left) and after (right) adjustment for potential half-periods in C24. Detections are flagged as potential half-periods (dark points) if, for at least one sector per target, the 2-term LS and ACF periods are both twice the LS period, within 10\% tolerance.}
    \label{fig:halfhalf}
\end{figure}

\begin{figure}
    \centering
    \includegraphics[width=0.8\textwidth]{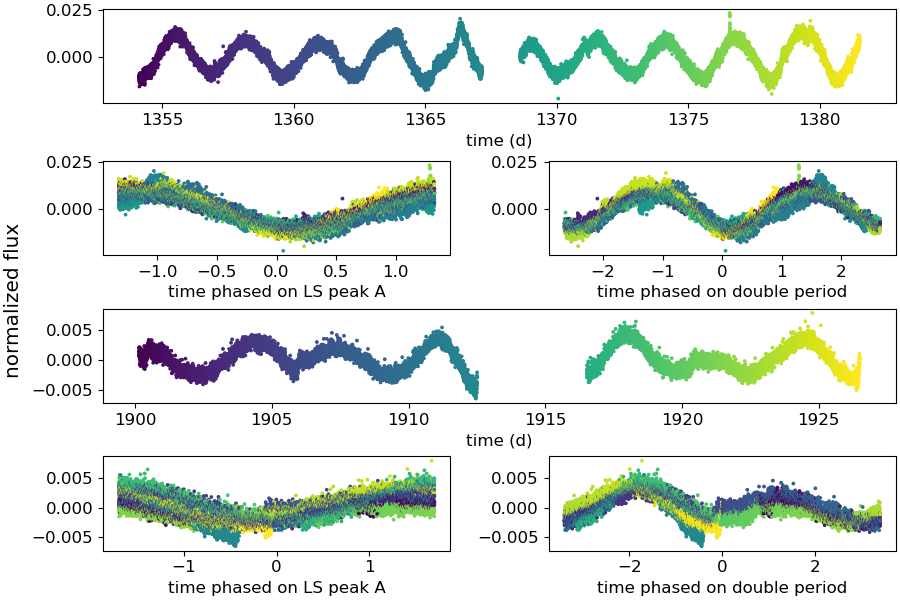}
    \caption{Two TESS light curves with points colored by time for comparison to folded versions, which are presented in the panels below. The left folded light curve is phased on the period detected by a LS periodogram in our pipeline, and the right is phased on twice that period. This allows us to visually identify double-dippers. For the first star (TIC 12359079, sector 2), the double-period phase curve has two peaks of equal flux, which suggests that the original period is in fact correct. For the second star (TIC 1126556, sector 22), the two peaks in the double-period curve are at different heights, which suggests that we are observing a double-dipping pattern. Indeed, the initial LS measurement for this star was flagged by our pipeline as being a likely half-period.}
    \label{fig:doubledip}
\end{figure}

Also notable in Figure~\ref{fig:xmatchres}, we appear to only observe half-periods (residuals~=~2), and not double-periods (residuals~=~0.5), of periods in both the O18 and H22 overlaps (though of course the LS failure modes occupy similar feature space). Reassuringly, we find no systematic dependence on \teff\ or TESS magnitude to explain this. The likely explanation for H22 and O18 periods over 10 days is that we have detected the half-period (or a higher harmonic frequency, in the case of very long periods in O18), but, in the regime where our pipeline is most applicable, it is also possible that detections in H22 and O18 may be double-periods. By select visual inspection of detections that lie along the half-period line with H22 (Figure~\ref{fig:halfhalf}, left panels), we note that in many cases, the LS period measurement appears correct. In the cases where it is evident that we have detected a half-period --- usually in a double-dipping rotation pattern \citep[e.g.][]{basri_double_2018} --- we find that the correct period is detected by both the 2-term LS periodogram and the ACF. This leads us to include a half-period flag and adjusted period for all targets that match this criterion within a 10\% tolerance range. Figure~\ref{fig:doubledip} compares a simple sinusoidal rotation pattern with a double-dipper, where in the latter case we have automatically detected that the LS period should be doubled to represent the correct period. The right-hand panels of Figure~\ref{fig:halfhalf} show the adjusted periods based on this metric in the H22 overlap sample. We find that a majority of periods along the half-period line now fall along the one-to-one line, and many periods with one-to-one correspondence are now flagged as double-periods when compared to H22. This distribution, with targets on both sides of the one-to-one line, is to be expected when crossmatching between two such catalogs.

\subsection{Discussion of methodologies} \label{sec:results4}

With the results of this analysis, we have the opportunity to directly compare the effectiveness of various approaches to retrieving rotation periods from TESS light curves. Our primary method for measuring rotation periods is the LS periodogram: we find that this is broadly the most indicative of rotation periods. In the case where a rotator exhibits a double-dipper pattern --- a complex periodic structure caused by multiple spots that manifests as two ``dips'' per cycle --- we find the 2-term LS periodogram and the ACF are both more robust for capturing the correct period. However, per \citet{basri_double_2018}, we expect to find fewer double-dippers in the period regime probed by this analysis, as double-dipper patterns tend mainly to emerge in longer-period rotators. This is consistent with the relatively low number of stars which we have flagged as half-periods: 1,796, or 16\% of the consolidated output catalog.

Due to the large overlap of targets detected by our pipeline and by \citet{holcomb_spinspotter_2022}, we are also able to compare the results of an ACF-based approach to our LS-based approach which uses the ACF as supplementary data. As can be seen in Figure~\ref{fig:catalogs}, we find that the LS has a harder upper limit than the ACF --- its flexibility as a pattern-matching tool is limited by the data downlink gap. However, this is less of an issue with the 2-term LS. As noted in both the training data and the final dataset, a combination of these approaches allows us to essentially push the detection limit for a number of high-amplitude, low-noise rotators. Adjusting for half-period detections in the single-term LS allows us to recreate the distribution found in \citet{holcomb_spinspotter_2022} (see Figure~\ref{fig:catalogs}). Conversely, the single-term LS seems to perform better for short periods, and the single-dip structure most common in this regime. This leads us to recommend a composite approach to detection in individual TESS sectors.

We can also comment on the use of the random forest algorithm as an approach to classification and vetting. Our approach eschews stellar parameters and trains only on light curve features, including data from three period-space measurements: with random forest classification, this has the effect of evenly distributing the results across input parameter space, such as \teff\ (see Figure~\ref{fig:pteff}). We find that the primary drawback of the random forest is that it requires a compromise between prioritizing true positives or true negatives: in order to limit the number of false positives, we recognize that some true positives are also lost. Our approach errs on the side of prioritizing the identification of true negatives (i.e. avoiding false positives), meaning that the final yield is not large, but it is high-fidelity. We were able to confirm this with select visual inspection as outlined in Section~\ref{sec:results1} ($>$12~day periods, the $\sim$6~day systematic), noting only a very small number of seeming false positives. In the majority of these cases, the pollutant was an eclipsing binary with an ellipsoidal component to its signal that emulates a rotational signal --- the actual eclipses were all truncated by outlier clipping as part of our data preparation. A statistically insignificant number of noisy light curves lacking any signal passed both classifiers --- these were all removed by the 0.01 (dimensionless) amplitude cut. Overall we find that, with adequate training, optimization, and post-processing, the random forest approach to vetting rotation in single TESS sectors is an effective way to determine both (1) the status of a signal as being caused by spot patterns and rotation, and (2) the accuracy of a period measurement.

\section{Conclusions} \label{sec:conclusions}

In this work we present a pipeline to detect and vet stellar rotation in individual TESS sectors, and an output catalog of periods detections across 10,909 targets. In the process of constructing the pipeline and analyzing its output, we have illuminated the upsides and cheallenges of this approach. Due to TESS systematics, it is hard to search for long periods, and stitching sectors presents its own issues. As such, we focused on individual sectors, which imposes a theoretical limit of $\sim$12~days due to the data downlink gap --- in practice, studies have found a sharp drop-off in detectability over $\sim$10~days \citep{kounkel_untangling_2022, holcomb_spinspotter_2022}. It is possible to surpass this limit with the pattern-matching capabilities of the LS periodogram and serendipitous signal timing. With visual inspection, we find 30 periods longer than 12~days that appear to be genuine detections of rotation.

Our approach when constructing this pipeline was to vet detections based solely on features that can be obtained from a light curve, so that temperature, magnitude, and all other stellar parameters are unknown to the random forests. Our training set was not necessarily complete in parameter space --- for example, there were no fast-rotating M dwarfs --- but we could nevertheless capture a detailed picture of the morphology of a light curve subject to spot modulation. The presence of fast-rotating M dwarfs in our final catalog confirms this. Our pipeline is open source and easy to implement, and of course there is yet more TESS data to be studied. Future work will involve processing the complete TESS dataset of FGKM stars in the extended mission, with a view to extending that work to untargeted FFI stars, potentially using custom photometry to robustly identify variability. The versatility of our pipeline also means that, given the absence of systematics, it can be used on longer time series of stitched sectors, such as \textit{TESS-SIP} \citep{hedges_systematics-insensitive_2020} or \textit{unpopular} \citep{hattori_unpopular_2021}.

In FGKM dwarfs from the first 26 sectors, we detected rotation in 3.75\% of all sectors analyzed, and 5.82\% of all targets. There are obvious reasons for this seemingly low figure: we face an upper limit on the period, impeding our ability to detect a large fraction of all rotators, as evident from the distributions retrieved from Kepler data by \citet{mcquillan_rotation_2014} and \citet{santos_surface_2021}, and in our training data \citep{oelkers_variability_2018}. On top of this, we apply a strict quality standard to our vetting procedure, optimizing for true negatives to minimize the presence of false positives in the final catalog. This has enabled us to provide a catalog of all 16,800 individual sector detections, and a consolidated catalog of 10,909 TESS targets, 4,936 of which are new. It is our hope that these high-fidelity rotation periods will be useful for empirical studies of gyrochronology, and that further studies using these detection and vetting methods will provide an expanded catalog data from the extended TESS mission and ongoing data releases.

\begin{acknowledgments}
Thanks to David Zurek and Sajesh Singh for their assistance with use of the computational facilities at the American Museum of Natural History. Thanks to Rae Holcomb and Zach Claytor for productive discussions about our varying approaches to the same task.

\end{acknowledgments}

%



\software{Astropy \citep{collaboration_astropy:_2013, collaboration_astropy_2018},
eleanor \citep{feinstein_eleanor_2019},
Lightkurve \citep{collaboration_lightkurve:_2021},
Matplotlib \citep{hunter_matplotlib_2007},
NumPy \citep{harris_array_2020},
Pandas \citep{reback_pandas_2021},
SciPy \citep{virtanen_scipy_2020},
GNU Parallel \citep{tange_ole_2018_1146014}
}




\bibliography{sample631}{}
\bibliographystyle{aasjournal}

\end{document}